\documentclass{article}
\usepackage{spconf,amsmath,graphicx}

\usepackage{enumitem}
\setlist{nosep, leftmargin=14pt}

\usepackage{mwe} % to get dummy images

\newcommand{\norm}[1]{\left\lVert#1\right\rVert}

\usepackage{caption}
\usepackage{subcaption}
\usepackage{graphicx}
\usepackage{hyperref}

\title{Continuous max-flow augmentation of self-supervised few-shot learning on SPECT left ventricles}
\twoauthors
 {Ádám István Szűcs} %\sthanks{Some author footnote.}
	{Eötvös Loránd University,\\
	Department of Computer Algebra,\\
	H-1117, Hungary, Budapest,\\ Pázmány P. blvd 1/C}
 {Béla Kári, Oszkár Pártos}
	{Semmelweis University,\\
	Department of Nuclear Medicine,\\
	HU-1082, Hungary, Budapest,\\ Üllői street 78a}
\begin{document}
%\ninept
%
\maketitle
\begin{abstract}
  Single-Photon Emission Computed Tomography (SPECT) left ventricular assessment protocols are important for detecting ischemia in high-risk patients. To quantitatively measure myocardial function, clinicians depend on commercially available solutions to segment and reorient the left ventricle (LV) for evaluation. Based on large normal datasets, the segmentation performance and the high price of these solutions can hinder the availability of reliable and precise localization of the LV delineation. To overcome the aforementioned shortcomings this paper aims to give a recipe for diagnostic centers as well as for clinics to automatically segment the myocardium based on small and low-quality labels on reconstructed SPECT, complete field-of-view (FOV) volumes. A combination of Continuous Max-Flow (CMF) with prior shape information is developed to augment the 3D U-Net self-supervised learning (SSL) approach on various geometries of SPECT apparatus. Experimental results on the acquired dataset have shown a 5-10\% increase in quantitative metrics based on the previous State-of-the-Art (SOTA) solutions, suggesting a good plausible way to tackle the few-shot SSL problem on high-noise SPECT cardiac datasets.
\end{abstract}
\section{Introduction}
\label{sec:intro}

Myocardial Perfusion Imaging (MPI) is aimed at getting results on hypertrophic cardiomyopathy and infarction with high specificity but at rather low sensitivity \cite{book}. This functional imaging technique with various radiopharmaceuticals is aimed at expressing metabolic tissue-level processes. As a side effect, implicitly purports the myocardial blood flow to discover ischemia in patients, since the administration of the pharmaceutical is injected intravenously. The MPI protocol is standardized by the European Association of Nuclear Medicine (EANM) and the American Society of Nuclear Cardiology (ASNC) by specifying the pharmaceutical administration as well as the stress or rest testing in different scenarios. However, with the wide range of imaging companies and equipment of SPECT cameras, assuring image quality remains a challenging task up to this day. Combining this high variability of acquisition with lowered doses in MPI, image processing becomes increasingly hard, especially segmentation.

To address and solve the aforementioned problems of myocardium segmentation in SPECT MPI, different machine learning algorithms have been developed \cite{garcia2021artificial, apostolopoulos2023deep}. Specifically convolutional neural networks (CNN) are one of the best candidates to segment the left ventricle \cite{hijazi2023deep}. Supervised training was conducted on cropped cardiac images with a 3D V-Net architecture by \cite{wang2020learning} utilizing binary cross entropy (BCE) with Dice similarity, resulting in a compound loss function. By taking the already great results further, Zhang et al. \cite{zhang2021novel} proposed a complete FOV-based spatial transformer network (STN) with 3D CNN to solve the segmentation and reorientation problem as well. Alongside the 3D CNN, another 2D U-Net-based network was proposed in \cite{wen2021analysis}, which is a good improvement over a dynamic programming (DP), and comparable performance was shown to the QGS software package \cite{Germano2007}. A different approach combining STN and CNN was used on whole image segmentation and reorientation, rendering great improvement over the latter approaches. And finally, the current SOTA results are conducted by \cite{ZHU2023106954}, combining 3D V-Net with shape priors on left ventricles, and in \cite{ni2023multi} multi-scale 3D U-Net with STN segmentation is used to segment complete FOV volumes with high performance. 

The literature is rich in the application and development of various learning strategies and architectures for the segmentation problem in complete FOV -- whole reconstructed volume -- segmentation, however, it becomes quite narrow, when one approaches the problem with small patient data at hand, namely with few-shot learning (FSL) or unsupervision. In this study small amount of labeled dataset ($<$10 patients) is considered with various conditions, where the predictions are tested on different imaging geometries and dosages on multiple patients.
\section{Methods}
\label{sec:methods}
In this paper, to segment the left ventricle on myocardial perfusion imaging SPECT, the complete reconstructed volume -- whole image -- is considered. The 3D volume of the scan is denoted $\Omega(x)$ as a closed domain. As the model of segmentation is thought of as continuous flow in three dimensions \cite{strang1983maximal,strang2009maximum} the source and sink terminal vertices are denoted as $s$ and $t$. At every $x\in \Omega$, $p(x)$ is the passing, $p_{s}(x)$ the directed flow from source to the current position, $p_{t}(x)$ is the directed sink flow from $x$ to $t$.

\subsection{3D U-Net SSL with augmentation}
\label{ssec:3d_unet_ssl}
The main backbone of the segmentation is 3D U-Net architecture \cite{cciccek20163d} with the following setup. The encoder part of the network consisted of three layers with Instance Normalization (IN) and PReLU activation functions. The various sized inputs were upsampled to $128^{3}$ sizes and were fed to the encoder part in $B \times C \times D \times H \times W$ order, B is batch, C is channel, H is height and W is the width. It should be noted that during the training of the encoder part of the network, the decoder was replaced by two dense layers with PReLU activation functions with Batch Normalization (BN) layer between with a dropout of $1/2$.

With very small labeled data at hand, the algorithm used heavy augmentation to reach the desired performance. For both stages of the self-supervised learning and supervised fine-tuning the augmentation pipeline consisted of the following processes, random blur, Gaussian noise, random affine, and elastic transformations with large parameter variations. These steps were applied in distinct modules in the pipeline and were obfuscated based on the best probabilities. However, after further testing, it was shown, that the network needed more augmentation to perform well. As in \cite{noroozi2016unsupervised}, to improve segmentation performance, the network had to solve jigsaw puzzles. The complete abdomen reconstructed 3D SPECT volumes were disassembled into tiles and randomly shuffled given as a task to solve for the network. The get the best out of this Representation Learning (RL) technique a large number of permutations, 1000 were used to perform the self-supervised pretext task. From the precomputed permutations the network had to choose the ones with maximal Hamming distances \cite{noroozi2016unsupervised}.

After the training of the encoder, it was loaded into a new network and the decoder was optimized by supervised-fine tuning on the 10 patient labeled dataset. The model utilized AdamW \cite{loshchilov2017decoupled}, with LogCosh loss function and a learning rate of \texttt{1e-3} with decay by a factor of $0.01$.

\subsection{Continuous Max-Flow enhancement}
\label{ssec:cmf}
The CNN 3D U-Net SSL \cite{adam2023} will generate a prediction map, the probability of each pixel belonging the the myocardium. This map is denoted by $\lambda(x),\ x \in \Omega$, where $\lambda(x) \in [0, 1]$ and $\lambda \in BV(\Omega)$, where $BV$ denotes the space of functions with bounded variation. This renders the original Mumford-Shah model \cite{mumford1989optimal} convex based on the Truncation lemma. The maximum flow problem is the dual of the minimal cut problem, which can be formalized as the following

\begin{equation}
  \label{eq:cmf}
  \begin{split}
      L_{c}(p_{s}, p_{t}, p, \lambda) &= \int_{\Omega} p_{s} dx \\ &+ \int_{\Omega} \lambda(\nabla \cdot p - p_{s} + p_{t}) dx \\ &- \frac{c}{2}\norm{\nabla \cdot p - p_{s} - p_{t}},
  \end{split}
\end{equation}
where $c>0$ and the solution is using the method by \cite{yuan2010study} and \cite{chambolle2004algorithm} as 
\begin{equation}
  \label{eq:tv_minimization}
  p^{k+1} = \Pi_{\alpha}\left( p^{k} + c \nabla( \nabla \cdot p^{k} - F^{k} ) \right),
\end{equation} 
where $\Pi_{\alpha}$ is a projection onto the convex set $C_{\alpha} =\{q\ |\ \norm{q} \leq \alpha\}$.

During the investigations on the test set, the Signal-to-Noise-Ratio rendered so low on different $^{201}$TI Chloride and some $^{99m}$Tc MIBI studies that the optimization got stuck in a local optima. To increase the performance of the scheme, a shape prior is introduced in the CMF method, with the aid of density estimation techniques.  

\subsection{Shape prior incorporation with density estimation}
\label{ssec:shape_prior}
Applying the 3D U-Net SSL approach on relatively normal patients showed good results in the evaluation. However, on hypoperfused myocardium, the approach was suboptimal as it was measured. Even increasing the dataset during the self-supervised pretext task the network was unable to segment dilated hearts.

To overcome these challenges, the Continuous Max-Flow enhancement incorporated prior information about the cardiac model. Since SPECT-based functional imaging varies a lot even in the same patient, with identical pharmacons at different times of the day, the analytical approach of the cardiac model is challenging. To relax the task of cardiac modeling of functional imaging, statistical evidence can help to build prior information about the different left ventricles. 

A similar approach is taken as in \cite{cremers2002diffusion}, where the outlined training data is encoded as a Gaussian distribution in higher dimensions. Given the negative logarithm of the probability, the Mahalanobis-type distance-based shape loss can be incorporated. Let $z$ be points in the higher dimensional space, then
\begin{equation}
  -\log{P(z)} = \frac{1}{2}(z - \mu)^{t} \Sigma^{-1}_{\perp} \frac{1}{2}(z - \mu),
\end{equation} 
where $\mu$ is the mean shape and $\Sigma^{-1}_{\perp}$ is the inverse of the regularized covariance matrix as in \cite{cremers2002diffusion}. This distance can act as a cost function in a probabilistic setting.

The problem with the latter approach is the assumption of the Gaussian approximation of the training shapes, which renders it obsolete for SPECT segmentation. To handle the non-gaussian nature and the complicated shape deformations of the training surfaces, the proposed model utilizes kernel space techniques \cite{cremers2003shape}, to give a good density estimation on prior variation. 
\section{Materials}
\label{sec:materials}

\subsection{Data}
\label{ssec:data}
Simulated data consisted of mathematical phantoms\cite{xcat}, whereas patient data consisted of 102 patient cases with different myocardial states in the perfusion protocol (non-gated) were evaluated. We created 3 different groups for various tests for our automation. (A) Perfusion studies with 53 patients, 23 females, and 30 males with an average age of 68.2 years, a body weight of 75.23, and a height of 164 cm. (B) Attenuation corrected perfusion studies consisted of 51 patients, 35 females, 16 males with an average age of 63.92 years, a body weight of 79.31 kgs and height of 166.62 cm. (C) Motion artifact perfusion studies of 36 patients, 19 female, 17 male with an average age of 53.97 years, the weight of 81 kgs, and height of 165.3 cm.

\subsection{Acquisition}
\label{ssec:acquisition}
Patient data acquisition was performed on different versions of Mediso AnyScan, AnyScan Trio SPECT, and SPECT/CT systems. The CT Image scanning parameters: SPIRAL acquisition type; CT Dose Index Volume 0.22; tube voltage 150 kVp; and tube current 30-50 mAs. Myocardial Perfusion SPECT (MPS) studies were performed at 4 participating sites with various isotopes $^{99m}$Tc MIBI, $^{99m}$Tc Tetrofosmin and $^{201}$TI Chloride. Acquisitions were performed with both 64 and 128 matrices, with 64 projections on a two-headed camera and 96 projections on a triple-headed camera. The angular interval was 180 on both dual and triple-headed cameras. Studies with attenuation maps were reconstructed with the TeraTomo method. Samples without attenuation correction (AC) utilized OSEM reconstruction method \cite{hudson1994accelerated} to create the complete FOV volume stack.

\section{Results}
\label{sec:results}
\begin{table*}[h!]
  \centering
  \begin{tabular}{ |p{2.3cm}|p{2.3cm}|p{2.3cm}|p{2.3cm}|p{2.3cm}|}
          \hline
          \multicolumn{5}{|c|}{Averaged performance metrics} \\
          \hline
           & Precision & Recall & IoU & Dice score \\
          \hline
          Proposed model & \textbf{0.7034} & 0.7286 & \textbf{0.5423} & \textbf{0.7065} \\
          \hline
          3D U-Net SSL & 0.5558 & \textbf{0.8551} & 0.5079 & 0.6737 \\
          \hline
          Zhu et. al & 0.0393 & 0.8456 & 0.0385 & 0.0725 \\
          \hline
  \end{tabular}
  \caption{Averaged segmentation results on the patient dataset. Our approach is capable of outperforming \cite{ZHU2023106954} and 3D U-Net SSL \cite{adam2023} as well on most of the metrics.}
  \label{tab:segmentation_results}
\end{table*}

\subsection{Quantitative evaluation}
\label{ssec:quant_ev}
The segmentation accuracy was evaluated on the following metrics. Accuracy = $\frac{TP + TN}{TP + TN + FP + FN}$, Precision = $\frac{TP}{TP + FP}$, Recall = $\frac{TP}{TP + FN}$, Intersection over Union (IoU) = $\frac{TP}{TP + FN + FP}$, DICE coefficient = $\frac{2 TP}{2TP + FP + FN}$, where T stands for true, F false, N is negative, P is positive.

\subsection{Implementation.} 
\label{ssec:implementation}
The code\footnote{\url{https://github.com/JacksonFurrier/isbi_2024_code.git}} is written in PyTorch \cite{NEURIPS2019_9015} and models were trained on a single Nvidia A100 GPU \cite{heder2022past} and a 1080Ti GPU as well. The annotated data consisted of 10 patients and the complete dataset contained 102 patients reconstructed complete FOV volumes. The implementation uses the TorchIO \cite{garcia2021artificial}, Scikit-learn \cite{scikit-learn} package, and extensively the GeomLoss \cite{feydy2019interpolating} library for the Sinkhorn loss and differentiation of cost functions.

\subsection{Experiments} 
\label{ssec:experiments}

As seen in Table \ref{tab:segmentation_results} the combined method of 3D SSL U-Net with CMF Shape prior enhancement reached the best results in segmenting the left ventricular volume of the complete-FOV reconstructed images. As the metric shows, the recall score is relatively higher, which lead to the conjecture that the method is over-predicting cardiac wall thickness. This problem can be two-fold, one possibility that the training data, along with the model of the left ventricle assumes thicker walls or the reconstruction method from kernel space lacks precision. This will need further tests and investigations, with combined oracle testing with MRI.

\begin{figure}[h!]
  \centering
  \begin{subfigure}[b]{0.22\textwidth}
  \centering
    \includegraphics[scale=0.13]{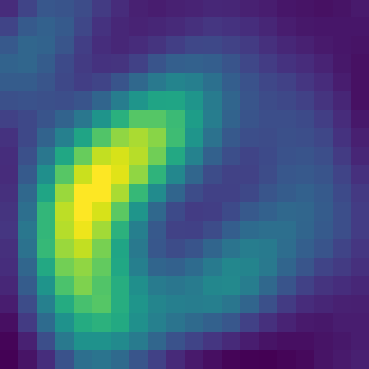}
    \includegraphics[scale=0.13]{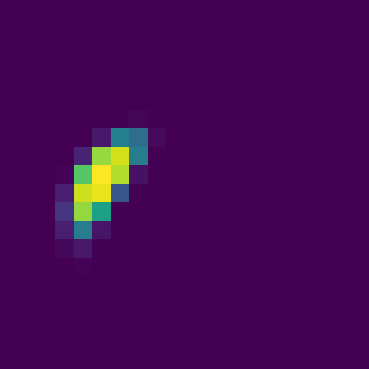}
    \includegraphics[scale=0.13]{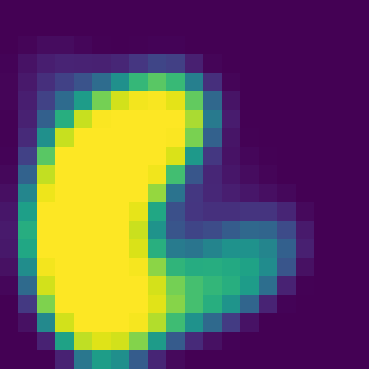}\\

    \includegraphics[scale=0.13]{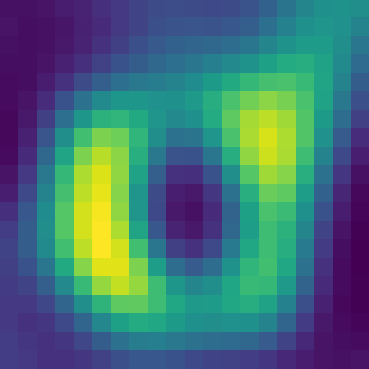}
    \includegraphics[scale=0.13]{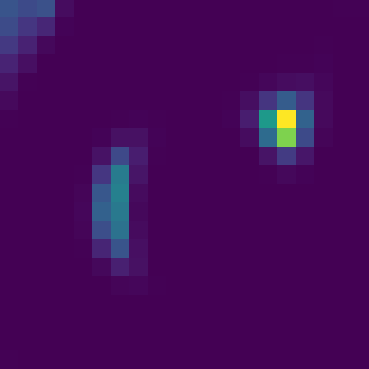}
    \includegraphics[scale=0.13]{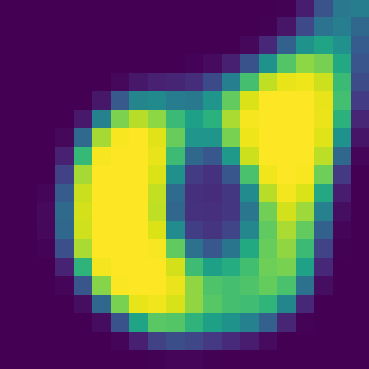}\\

    \includegraphics[scale=0.13]{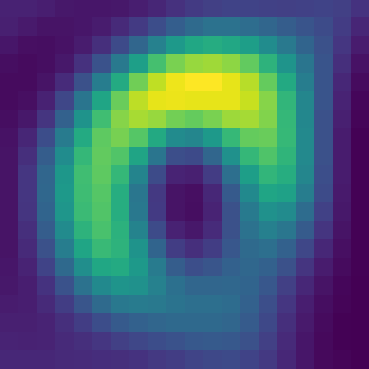}
    \includegraphics[scale=0.13]{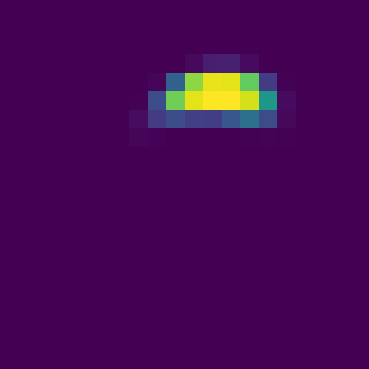}
    \includegraphics[scale=0.13]{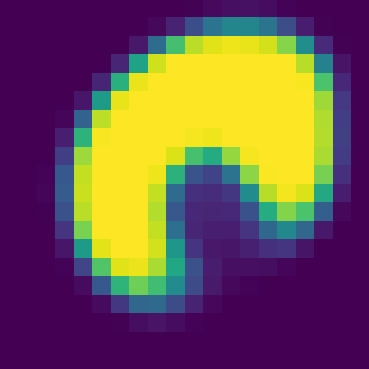}\\
    \caption{}
    \label{fig:pat_tc99m_stable_perfusion_defect}
  \end{subfigure}
  \begin{subfigure}[b]{0.22\textwidth}
    \centering
      \includegraphics[scale=0.13]{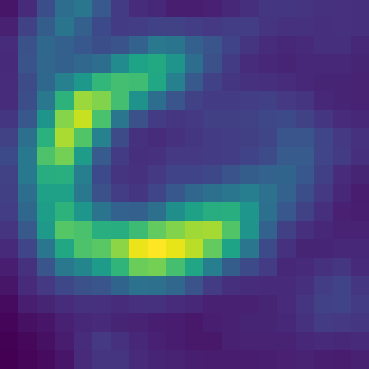}
      \includegraphics[scale=0.13]{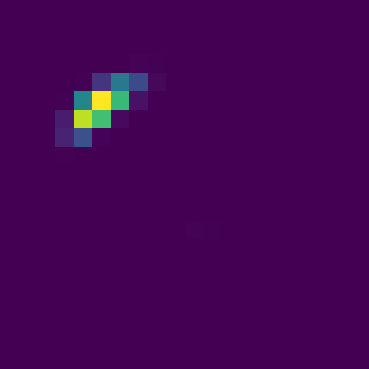}
      \includegraphics[scale=0.13]{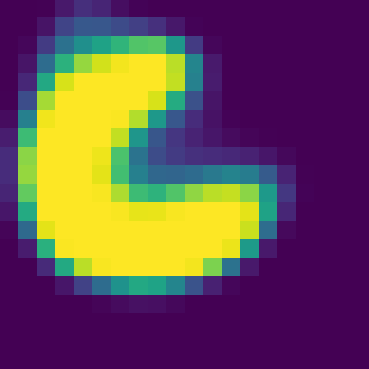}\\
  
      \includegraphics[scale=0.13]{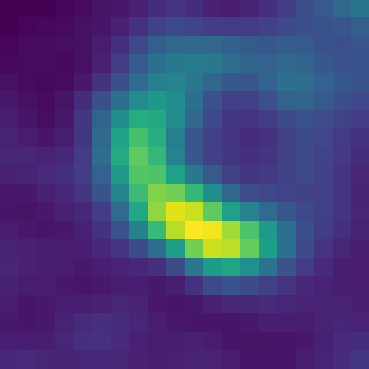}
      \includegraphics[scale=0.13]{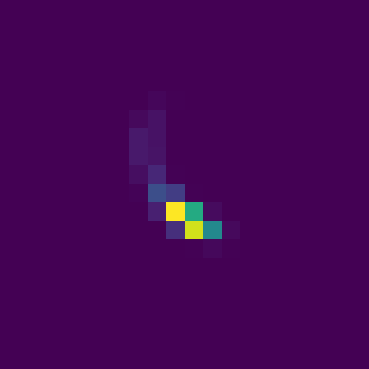}
      \includegraphics[scale=0.13]{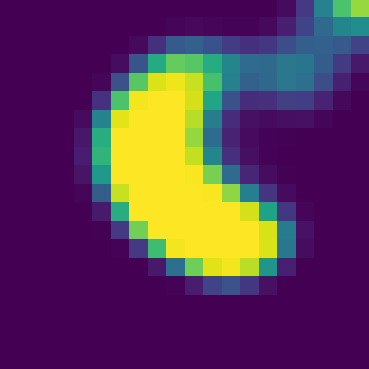}\\
  
      \includegraphics[scale=0.13]{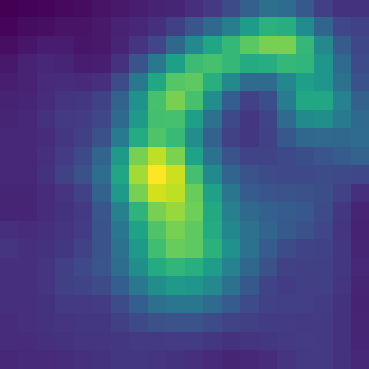}
      \includegraphics[scale=0.13]{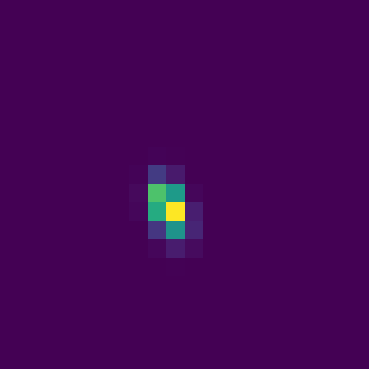}
      \includegraphics[scale=0.13]{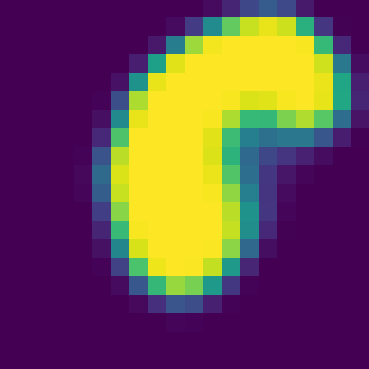}\\
      \caption{}
      \label{fig:pat_tc99_inferior_pref_defect}
    \end{subfigure}

  \caption{Enhancement of the 3D SSL U-Net \cite{adam2023} approach with CMF Shape prior segmentation. The first column depicts the SPECT volume of the cardiac area, the second column shows the segmentation with 3D SSL U-Net, and the third column shows the results with CMF shape prior enhancement. Two patients with different cardiac conditions, \ref{fig:pat_tc99m_stable_perfusion_defect} with stable perfusion defect and \ref{fig:pat_tc99_inferior_pref_defect} patient showing signs of inferior perfusion defect. }
  \label{fig:defected_myocardium_results}
\end{figure}

Knowing that the 3D SSL U-Net performance was deteriorating in hypoperfused cardiac patients, the tests consisted of evaluation with ill-conditioned patients. As seen on \ref{fig:defected_myocardium_results} the CMF enhancement outperformed the original 3D SSL U-Net approach, however, as it was previously mentioned the ``thickening'' of the cardiac wall persists.

\section{Conclusion}
\label{ssec:conclusion}
In this study, a different approach is taken in ``augmenting''
the 3D U-Net SSL technique for segmenting normal and heavily hypoperfused SPECT left ventricles. The contribution lies in the combination of 3D SSL U-Net with a Shape prior-based Continuous Max-Flow optimization. The results show that further investigation of shape prior information involvement in deep learning segmentation tasks could be a viable approach to overcome the issues of a few labeled datasets.

\section{Compliance with ethical standards}
\label{sec:ethics}

\begin{itemize}
    \item ``This study was performed in line with the principles of
      the Declaration of Helsinki. Approval was granted by the Ethics
      Committee of Semmelweis University''
\end{itemize}

\section{Acknowledgments}
\label{sec:acknowledgments}

The research was supported by the project No. 2019-1.3.1-KK-2019-00011 financed by the National Research, Development and Innovation Fund of Hungary under the Establishment of Competence Centers, Development of Research Infrastructure Programme funding scheme. Further, on behalf of the MedisoLab project, we are grateful for the possibility of using ELKH Cloud (see Héder et al. 2022 \cite{heder2022past}; https://science-cloud.hu/) which helped us achieve the results published in this paper.

\bibliographystyle{IEEEbib}
\bibliography{refs}

\begin{thebibliography}{10}

\bibitem{book}
M~{Elizabeth Oates} and Vincent Sorrell,
\newblock {\em {Myocardial Perfusion Imaging - Beyond the Left Ventricle}},
\newblock Springer International Publishing, 2017.

\bibitem{garcia2021artificial}
Ernest~V Garcia,
\newblock ``Artificial intelligence in nuclear cardiology: Preparing for the
  fifth industrial revolution,''
\newblock {\em Journal of Nuclear Cardiology}, vol. 28, pp. 1199--1202, 2021.

\bibitem{apostolopoulos2023deep}
Ioannis~D Apostolopoulos, Nikolaos~I Papandrianos, Anna Feleki, Serafeim
  Moustakidis, and Elpiniki~I Papageorgiou,
\newblock ``Deep learning-enhanced nuclear medicine spect imaging applied to
  cardiac studies,''
\newblock {\em EJNMMI physics}, vol. 10, no. 1, pp. 6, 2023.

\bibitem{hijazi2023deep}
Waseem Hijazi and Robert~JH Miller,
\newblock ``Deep learning to automate spect mpi myocardial reorientation,''
\newblock {\em Journal of Nuclear Cardiology}, pp. 1--2, 2023.

\bibitem{wang2020learning}
Tonghe Wang, Yang Lei, Haipeng Tang, Zhuo He, Richard Castillo, Cheng Wang,
  Dianfu Li, Kristin Higgins, Tian Liu, Walter~J Curran, et~al.,
\newblock ``A learning-based automatic segmentation and quantification method
  on left ventricle in gated myocardial perfusion spect imaging: A feasibility
  study,''
\newblock {\em Journal of Nuclear Cardiology}, vol. 27, pp. 976--987, 2020.

\bibitem{zhang2021novel}
Duo Zhang, P~Hendrik Pretorius, Kaixian Lin, Weibing Miao, Jingsong Li,
  Michael~A King, and Wentao Zhu,
\newblock ``A novel deep-learning--based approach for automatic reorientation
  of 3d cardiac spect images,''
\newblock {\em European Journal of Nuclear Medicine and Molecular Imaging},
  vol. 48, pp. 3457--3468, 2021.

\bibitem{wen2021analysis}
Haixing Wen, Qiuyue Wei, Jin-Long Huang, Shih-Chuan Tsai, Chi-Yen Wang,
  Kuo-Feng Chiang, Yu~Deng, Xiongtao Cui, Rui Gao, Weihua Zhou, et~al.,
\newblock ``Analysis on spect myocardial perfusion imaging with a tool derived
  from dynamic programming to deep learning,''
\newblock {\em Optik}, vol. 240, pp. 166842, 2021.

\bibitem{Germano2007}
Guido Germano, Paul~B. Kavanagh, Piotr~J. Slomka, Serge~D. {Van Kriekinge},
  Geoff Pollard, and Daniel~S. Berman,
\newblock ``{Quantitation in gated perfusion SPECT imaging: The Cedars-Sinai
  approach},''
\newblock {\em Journal of Nuclear Cardiology}, vol. 14, no. 4 SPEC. ISS., pp.
  433--454, 2007.

\bibitem{ZHU2023106954}
Fubao Zhu, Longxi Li, Jinyu Zhao, Chen Zhao, Shaojie Tang, Jiaofen Nan, Yanting
  Li, Zhongqiang Zhao, Jianzhou Shi, Zenghong Chen, Chuang Han, Zhixin Jiang,
  and Weihua Zhou,
\newblock ``A new method incorporating deep learning with shape priors for left
  ventricular segmentation in myocardial perfusion spect images,''
\newblock {\em Computers in Biology and Medicine}, vol. 160, pp. 106954, 2023.

\bibitem{ni2023multi}
Yangfan Ni, Duo Zhang, Gege Ma, Lijun Lu, Zhongke Huang, and Wentao Zhu,
\newblock ``A multi-scale spatial transformer u-net for simultaneously
  automatic reorientation and segmentation of 3d nuclear cardiac images,''
\newblock {\em arXiv preprint arXiv:2310.10095}, 2023.

\bibitem{strang1983maximal}
Gilbert Strang,
\newblock ``Maximal flow through a domain,''
\newblock {\em Mathematical Programming}, vol. 26, pp. 123--143, 1983.

\bibitem{strang2009maximum}
Gilbert Strang,
\newblock ``Maximum flows and minimum cuts in the plane,''
\newblock {\em Advances in Applied Mathematics and Global Optimization: In
  Honor of Gilbert Strang}, pp. 1--11, 2009.

\bibitem{cciccek20163d}
{\"O}zg{\"u}n {\c{C}}i{\c{c}}ek, Ahmed Abdulkadir, Soeren~S Lienkamp, Thomas
  Brox, and Olaf Ronneberger,
\newblock ``3d u-net: learning dense volumetric segmentation from sparse
  annotation,''
\newblock in {\em Medical Image Computing and Computer-Assisted
  Intervention--MICCAI 2016: 19th International Conference, Athens, Greece,
  October 17-21, 2016, Proceedings, Part II 19}. Springer, 2016, pp. 424--432.

\bibitem{noroozi2016unsupervised}
Mehdi Noroozi and Paolo Favaro,
\newblock ``Unsupervised learning of visual representations by solving jigsaw
  puzzles,''
\newblock in {\em Computer Vision--ECCV 2016: 14th European Conference,
  Amsterdam, The Netherlands, October 11-14, 2016, Proceedings, Part VI}.
  Springer, 2016, pp. 69--84.

\bibitem{loshchilov2017decoupled}
Ilya Loshchilov and Frank Hutter,
\newblock ``Decoupled weight decay regularization,''
\newblock {\em arXiv preprint arXiv:1711.05101}, 2017.

\bibitem{adam2023}
Adam Istvan~Szucs et~al.,
\newblock ``Self-supervised segmentation of myocardial perfusion imaging spect
  left ventricles,''
\newblock in {\em 2023 the 10th International Conference on Bioinformatics
  Research and Applications (ICBRA) (ICBRA 2023), September 22-24, 2023,
  Barcelona, Spain.} ACM, 2023.

\bibitem{mumford1989optimal}
David~Bryant Mumford and Jayant Shah,
\newblock ``Optimal approximations by piecewise smooth functions and associated
  variational problems,''
\newblock {\em Communications on pure and applied mathematics}, 1989.

\bibitem{yuan2010study}
Jing Yuan, Egil Bae, and Xue-Cheng Tai,
\newblock ``A study on continuous max-flow and min-cut approaches,''
\newblock in {\em 2010 ieee computer society conference on computer vision and
  pattern recognition}. IEEE, 2010, pp. 2217--2224.

\bibitem{chambolle2004algorithm}
Antonin Chambolle,
\newblock ``An algorithm for total variation minimization and applications,''
\newblock {\em Journal of Mathematical imaging and vision}, vol. 20, pp.
  89--97, 2004.

\bibitem{cremers2002diffusion}
Daniel Cremers, Florian Tischh{\"a}user, Joachim Weickert, and Christoph
  Schn{\"o}rr,
\newblock ``Diffusion snakes: Introducing statistical shape knowledge into the
  mumford-shah functional,''
\newblock {\em International journal of computer vision}, vol. 50, pp.
  295--313, 2002.

\bibitem{cremers2003shape}
Daniel Cremers, Timo Kohlberger, and Christoph Schn{\"o}rr,
\newblock ``Shape statistics in kernel space for variational image
  segmentation,''
\newblock {\em Pattern Recognition}, vol. 36, no. 9, pp. 1929--1943, 2003.

\bibitem{xcat}
W.~P. Segars, G.~Sturgeon, S.~Mendonca, Jason Grimes, and B.~M.~W. Tsui,
\newblock ``4d xcat phantom for multimodality imaging research,''
\newblock {\em Medical Physics}, vol. 37, no. 9, pp. 4902--4915, 2010.

\bibitem{hudson1994accelerated}
H~Malcolm Hudson and Richard~S Larkin,
\newblock ``Accelerated image reconstruction using ordered subsets of
  projection data,''
\newblock {\em IEEE transactions on medical imaging}, vol. 13, no. 4, pp.
  601--609, 1994.

\bibitem{NEURIPS2019_9015}
Adam Paszke and et~al.,
\newblock ``Pytorch: An imperative style, high-performance deep learning
  library,''
\newblock pp. 8024--8035. Curran Associates, Inc., 2019.

\bibitem{heder2022past}
Mih{\'a}ly H{\'e}der, Ern{\H{o}} Rig{\'o}, Dorottya Medgyesi, R{\'o}bert Lovas,
  Szabolcs Tenczer, Attila Farkas, M{\'a}rk~Benj{\'a}min Em{\H{o}}di,
  J{\'o}zsef Kadlecsik, and P{\'e}ter Kacsuk,
\newblock ``The past, present and future of the elkh cloud,''
\newblock {\em INFORM{\'A}CI{\'O}S T{\'A}RSADALOM: T{\'A}RSADALOMTUDOM{\'A}NYI
  FOLY{\'O}IRAT}, vol. 22, no. 2, pp. 128--137, 2022.

\bibitem{scikit-learn}
F.~Pedregosa, G.~Varoquaux, A.~Gramfort, V.~Michel, B.~Thirion, O.~Grisel,
  M.~Blondel, P.~Prettenhofer, R.~Weiss, V.~Dubourg, J.~Vanderplas, A.~Passos,
  D.~Cournapeau, M.~Brucher, M.~Perrot, and E.~Duchesnay,
\newblock ``Scikit-learn: Machine learning in {P}ython,''
\newblock {\em Journal of Machine Learning Research}, vol. 12, pp. 2825--2830,
  2011.

\bibitem{feydy2019interpolating}
Jean Feydy, Thibault S{\'e}journ{\'e}, Fran{\c{c}}ois-Xavier Vialard, Shun-ichi
  Amari, Alain Trouve, and Gabriel Peyr{\'e},
\newblock ``Interpolating between optimal transport and mmd using sinkhorn
  divergences,''
\newblock in {\em The 22nd International Conference on Artificial Intelligence
  and Statistics}, 2019, pp. 2681--2690.

\end{thebibliography}

\end{document}